%% file: magnification_bias.tex
\documentclass[fleqn,usenatbib]{mnras}
\usepackage{newtxtext,newtxmath}

\usepackage[T1]{fontenc}
\usepackage{ae,aecompl}

\usepackage{graphicx}	
\usepackage{amsmath}	
\usepackage{color}
\usepackage[dvipsnames]{xcolor}
\usepackage{booktabs,threeparttable}
\usepackage{soul}
\usepackage{siunitx}

\usepackage{tikz}
\usetikzlibrary{shapes,arrows}

\usepackage{colortbl}
\usepackage{physics}

\input{journaldef.tex}

\newcommand{\bea}{\begin{eqnarray}}
\newcommand{\be}{\begin{equation}}
\newcommand{\ben}{\begin{enumerate}}
\newcommand{\bi}{\begin{itemize}}
\newcommand{\eea}{\end{eqnarray}}
\newcommand{\ee}{\end{equation}}
\newcommand{\ei}{\end{itemize}}
\newcommand{\een}{\end{enumerate}}

\usepackage{xspace}

\renewcommand{\P}{\mathbf P}

\newcommand{\ccl}{\textsc{CCL}\xspace}

\definecolor{azure}{rgb}{0.0, 0.5, 1.0}
\definecolor{darkgreen}{cmyk}{0.85,0.2,1.00,0.2} 

\usepackage{siunitx}
\DeclareSIUnit \h {\ensuremath{\mathit{h}}}
\DeclareSIUnit \Msun {M_\odot}
\DeclareSIUnit \parsec {pc}
\DeclareSIUnit \deg {deg}

\usepackage[capitalise]{cleveref}

\title[Magnification Bias Estimators for Realistic Surveys]{Magnification Bias Estimators for Realistic Surveys: an Application to the BOSS Survey}

\author[Wenzl et al.]{Lukas Wenzl$^{1}$\thanks{E-mail:ljw232@cornell.edu}; Shi-Fan Chen$^{2}$; Rachel Bean$^{1}$
\\
$^{1}$ Department of Astronomy, Cornell University, Ithaca, NY, 14853, USA \\
$^{2}$ Institute for Advanced Study, 1 Einstein Drive, Princeton, NJ 08540, USA \\
}
\date{Accepted 2023 October 23. Received 2023 October 19; in original form 2023 August 10}

\pubyear{2023}

\begin{document}
\label{firstpage}
\pagerange{\pageref{firstpage}--\pageref{lastpage}}
\maketitle

\begin{abstract} 
In addition to the intrinsic clustering of galaxies themselves, the spatial distribution of galaxies observed in surveys is modulated by the presence of weak lensing due to matter in the foreground. This effect, known as magnification bias, is a significant contaminant to analyses of galaxy-lensing cross-correlations and must be carefully modelled. We present a method to estimate the magnification bias in spectroscopically confirmed galaxy samples based on finite differences of galaxy catalogues while marginalizing over errors due to finite step size. We use our estimator to measure the magnification biases of the CMASS and LOWZ samples in the SDSS BOSS galaxy survey, analytically taking into account the dependence on galaxy shape for fiber and PSF magnitudes, finding $\alpha_{\rm CMASS} = 2.71 \pm 0.02$ and $\alpha_{\rm LOWZ} = 2.45 \pm 0.02$ and quantify modelling uncertainties in these measurements. Finally, we quantify the redshift evolution of the magnification bias within the CMASS and LOWZ samples, finding a difference of up to a factor of three between the lower and upper redshift bounds for the former. We discuss how to account for this evolution in modelling and its interaction with commonly applied redshift-dependent weights. Our method should be readily-applicable to upcoming surveys and we make our code publicly available as part of this work. 
\end{abstract}


\section{Introduction}
\label{sec:intro}

Recent years have seen rapid advances in joint analyses of galaxy clustering and weak lensing surveys, which have become indispensable tools in measuring the late-time growth of structure \citep{Abbott2018,Heymans2020,Krolewski21,Miyatake2022}, especially as compared to the $\Lambda$CDM models preferred by cosmic microwave background (CMB) measurements \citep[see e.g.][]{Planck}. 

Gravitational lensing results from the deflection of photons from distant sources by the gravitational potentials of intervening matter along the line of sight \citep[see e.g.][]{Bartelmann2001}. In galaxy surveys, in the weak gravitational limit, this leads to small shear distortions in the shapes of distant galaxies that are inherently correlated with the clustering of foreground matter and galaxies. In CMB surveys, lensing by galaxies coherently remaps the CMB primary anisotropies to different angular modes \citep{Zaldarriaga:1998te,Hu:2000ax,Lewis:2006fu}. 

The cross-correlation signals of foreground lensing galaxies and background ones (galaxy-galaxy lensing) and CMB anisotropies (galaxy-CMB lensing) are sensitive to the amplitude of matter clustering at the epoch wherein the galaxies reside, allowing us to probe the history of cosmic growth tomographically, i.e. at isolated redshift slices \citep[e.g.][]{Massey:2007gh,Heymans:2013fya,Mandelbaum_2013,DES:2015eqk,GarciaGarcia2021,White2022}. Moreover, comparing the strength of matter clustering, as probed through the deflections of photons by gravity, versus through the dynamics of massive tracers like galaxies, through redshift-space distortions in spectroscopic surveys at the same epoch, allows us to test the predictions of general relativity against theories of modified gravity \citep{Zhang:2007nk,Reyes:2010tr,Simpson:2012ra,Leonard:2015cba,Blake:2015vea,Pullen2015,delaTorre:2016rxm,Pullen2016,Alam:2016qcl,Amon:2017lia,Singh2019}. 

In addition to the shear distortions lensing also induces signal magnification \citep{Hildebrandt_2009,Schmidt:2011qj,Duncan:2013haa,Hildebrandt:2015kcb,Thiele:2019fcu, Unruh:2019pyk}. In addition to being a measurable cosmological signal in its own right \citep{Jain02,2005ApJ...633..589S,Schmidt12}, this magnification induces an important systematic in galaxy-lensing cross-correlations, as well as galaxy autocorrelations, known in the literature as magnification bias. The lensing contribution has a non-negligible impact for the current and upcoming generation of cosmology surveys \citep{Krolewski2020,Maartens21,vonWietersheimKramsta2021,Duncan22,Lepori22,ElvinPoole2023} and as such is important to properly model in analyses of the cross-correlations signal.

The cosmological constraining potential of galaxy-galaxy and galaxy-CMB correlations promises to be substantial with upcoming large-scale structure and CMB surveys of increased sky coverage, redshift depth and measurement precision. This includes spectroscopic surveys such as the Dark Energy Spectroscopic Instrument \citep[DESI\footnote{\url{https://www.desi.lbl.gov/}},][]{DESICollaboration2016}, and the Prime Focus Spectrograph \citep[PFS\footnote{\url{https://pfs.ipmu.jp/}},][]{Takada2014}, joint photometric and spectroscopic surveys, \textit{Euclid}\footnote{\url{https://sci.esa.int/web/euclid}} \citep{Laureijs2011} and the Nancy Grace Roman Space Telescope \citep[\textit{Roman}\footnote{\url{https://roman.gsfc.nasa.gov/}},][]{Spergel2015}, the Spectro-Photometer for the History of the Universe, Epoch of Reionization, and Ices Explorer \citep[SPHEREx\footnote{\url{http://spherex.caltech.edu/}},][]{Dore2014}  and also applies for photometric surveys like the Vera C. Rubin Observatory Legacy Survey of Space and Time \citep[LSST\footnote{\url{https://www.lsst.org/}},][]{Ivezic2019}. The significant increase in statistical power of these surveys makes it timely to investigate systematics biases in cosmological inference with these surveys, for example, due to approximate modelling choices.

The goal of this paper is to carefully formulate a procedure to calculate the magnitude bias precisely for a realistic galaxy sample, with spectroscopic samples in particular in mind. We demonstrate the implementation of the procedure, and perform a measurement of the effect, for galaxies in the Sloan Digital Sky Survey Baryon Oscillations Spectroscopic Survey \citep[SDSS BOSS][]{Dawson13}. In \cref{sec:SDSS_data}, we briefly describe the SDSS datasets to establish the photometric selection and observational properties leveraged for our estimation. In \cref{sec:mag_bias_derivation}, we lay out the method to determine the magnification bias for realistic spectroscopic surveys, resulting in generally applicable estimators. A key part of estimating the magnification bias is to accurately model the change of the observed magnitudes under lensing, which we carefully consider in \cref{sec:lensing_mags}. In \cref{sec:application_SDSS_data}, we present measurements of the magnification bias for the SDSS samples using our estimators. We draw together our conclusions and discuss implications for future work in section \cref{sec:conclusion}. Some additional technical details about redshift evolution are given in \cref{appendix:redshift_dependence}.

\section{SDSS BOSS Data}
\label{sec:SDSS_data}

\begin{figure}
\includegraphics[width=\columnwidth]{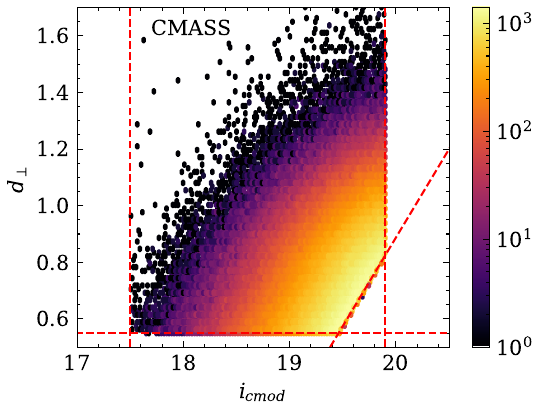}
\includegraphics[width=\columnwidth]{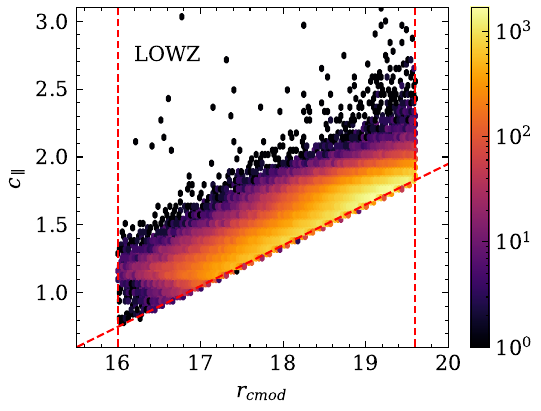}
\caption{Visualizing part of the photometric selection with cuts summarized in \cref{table:magnification_bias}. [Upper] Selection cuts 1, 2, and 3 for the CMASS sample and [Lower] cuts 1 and 2 for the LOWZ sample. The selections are not purely magnitude-limited. To estimate the magnification bias we need to account for the full photometric selection. \label{fig:color_selection_CMASS}}
\end{figure}

Throughout this paper, our focus will be on the magnification properties of galaxies in SDSS BOSS \citep{Dawson13}. BOSS was a redshift survey as part of Sloan Digital Sky Survey III \citep{SDSSIII}, covering 10,252 square degrees of sky and 1,198,006 galaxies. In particular, we will use the catalogues from the final data release (DR12) as described in \citet{Reid2016}; these are given as cuts on functions of variously-defined colours and magnitudes (e.g. ``model'', ``cModel'', PSF, and fiber2 magnitudes\footnote{We use the standard SDSS notation referring to them as $b_{\mathrm{mod}}, b_{\mathrm{cmod}},b_{\mathrm{psf}},b_{\mathrm{fib2}}$ for each band $b\in [u,g,r,i,z]$}) on imaging data from SDSS\footnote{We follow the procedure for the original selection, using the measured fluxes $f$ in the unit 'nanomaggy' and converting them to Pogson magnitudes as
\begin{equation}
    m = 22.5 - 2.5 \log_{10} \left( \textrm{max} (f, 0.001)\right).
\end{equation}
We note that we are specifically not using the Asinh Magnitudes provided in the SDSS catalogues. While they are more stable for low SNR and do not require clipping of negative fluxes, using them would differ from the original selection which leads to issues close to the boundary that would slightly bias our results. The code for the original selection can be found at \url{https://data.sdss.org/sas/dr12/boss/lss/mksampleDR12}.}, from which two spectroscopic (target) samples---LOWZ and CMASS---at low and high redshifts were selected. A subset of these cuts is shown in Fig.~\ref{fig:color_selection_CMASS}  \citep[see also Fig.~4 in][]{Reid2016}. Importantly, the most significant survey boundaries are not given by straightforward magnitude cuts (e.g. cuts in the i-band and r-band ``cModel'' magnitudes in the upper and lower panels) but rather by colour-dependent ($d_\perp$ and $c_\parallel$ in this case) magnitude cuts around which most galaxies cluster. Additionally, cuts using the fiber and point spread function-convolved (PSF) magnitudes were applied to the data. An overview of all photometric selection criteria for CMASS and LOWZ can be found in \cref{table:magnification_bias}. The additional colour combinations used in the selection are defined as $d_\perp = (r_{\mathrm{mod}}-i_{\mathrm{mod}}) - (g_{\mathrm{mod}}-r_{\mathrm{mod}})/8$,  $c_\perp = (r_{\mathrm{mod}}-i_{\mathrm{mod}})-(g_{\mathrm{mod}}-i_{\mathrm{mod}})/4 - 0.18$ and $c_\parallel = 0.7 (g_{\mathrm{mod}}-r_{\mathrm{mod}}) + 1.2 (r_{\mathrm{mod}}-i_{\mathrm{mod}} - 0.18)$.

From the baseline CMASS and LOWZ selections two distinct LSS-oriented sets of samples were created: the LSS LOWZ and CMASS samples, with additional redshift cuts applied at $0.15 < z < 0.43$ and $0.43 < z < 0.75$, respectively, and a merged sample combining LOWZ\footnote{Many of the early data chunks feature different photometric selections for LOWZ galaxies, as detailed in Appendix A of \citet{Reid2016}. These are not included in the fiducial LOWZ samples but were merged into the combined z1 and z3 LSS samples described next, and as such it is necessary to properly account for these early data chunks separately when dealing with these alternative catalogues which amounts to averaging over different galaxy samples across the sky. We include the additional cuts for the early data chunks in our analysis.} and CMASS with redshift cuts for low and high redshift bins z1 and z3 between $0.2 < z <0.5$ and $0.5 < z < 0.75$. We work exclusively with these galaxy catalogues in this paper.

Known systematics in the data require the use of weights to infer unbiased cosmological constraints with these catalogues. The weighting scheme 
\begin{equation}
w_g = (w_{\rm NOZ} + w_{\rm CP} -1)\cdot  w_{\rm SEEING} \cdot w_{\rm STAR}
\end{equation}
accounts for trends in seeing $w_{\rm SEEING}$, galactic latitude $w_{\rm STAR}$, redshift failures $w_{\rm NOZ}$, and close pairs $w_{\rm CP}$. We account for these weights throughout our analysis and they have important implications for the impact of the spectroscopic success rate (\cref{sec:spectroscopic_success_rate}). Additionally, we will consider the impact of commonly used additional weights: FKP weights \citep{Feldman:1993ky,Pearson:2016jzc} as well as weights to match the effective redshift of cross-correlations to auto-correlations \citep[see e.g.][]{Chen2022}.

\section{Magnification Bias in Spectroscopic Surveys} \label{sec:mag_bias_derivation}

\subsection{Background on magnification bias}
\label{sec:magbiasoverview}

\begin{figure}
\includegraphics[width=\columnwidth]{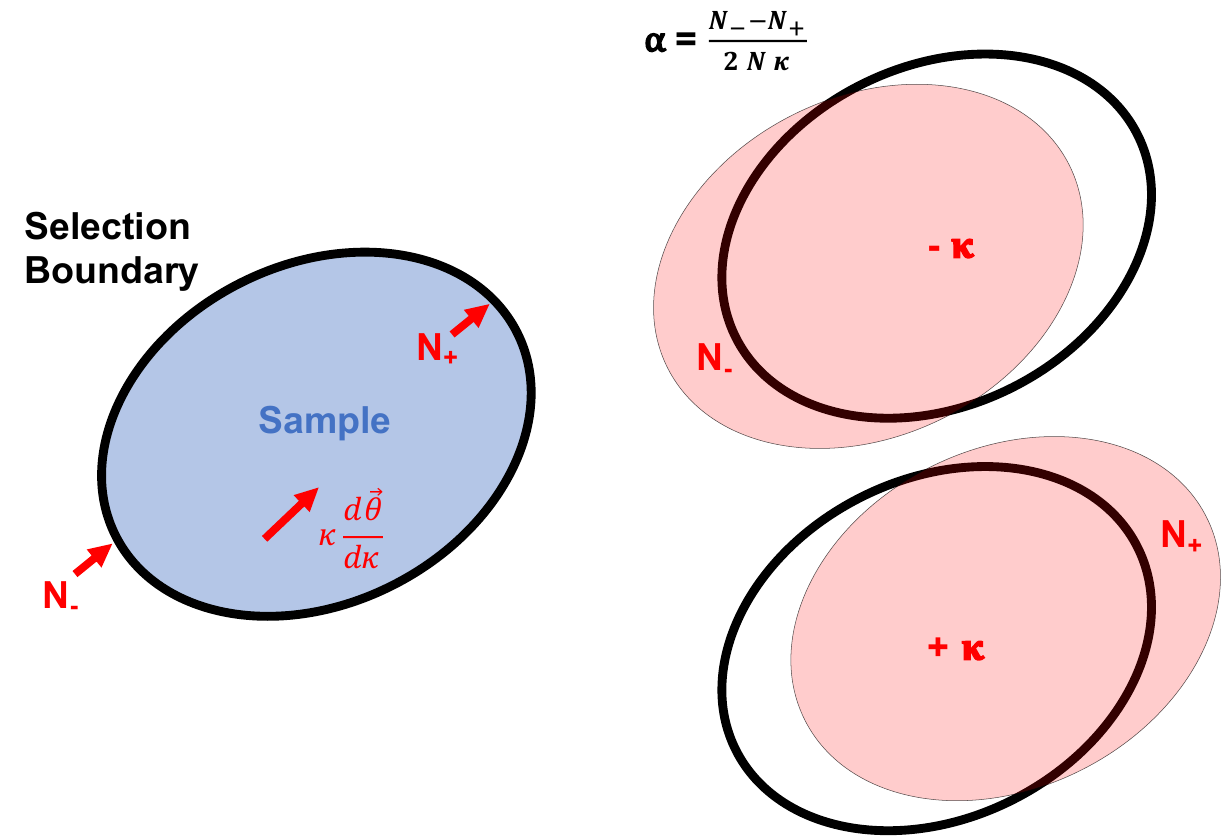}
\caption{Schematic of the magnification bias effect. [Left] For a given survey selection, shown here as a black oval, an added lensing convergence $\kappa$ induces a flow of the selected parameters $\lambda(\kappa)$ into and out of the selected parameter space. [Right] If the amount that flows out of (into) the selection boundaries are denoted by $N_{\pm}$, then the magnification bias $\alpha$ can be estimated using a simple finite difference.
\label{fig:mag_graphic}}
\end{figure}

The magnification bias effect was first described for magnitude-limited surveys (see e.g. \cite{Bartelmann2001}) where all galaxies in a given sample with magnitudes below $m_{\rm cut}$ are selected. In this case, applying a foreground magnification $\mu = 1 + 2\kappa$, where $\kappa$ is the lensing convergence, changes the observed number density of galaxies by
\begin{align}
    \frac{ \Delta n}{n} &= (1 - 2\kappa)\ n(< m_{\rm cut} + 5 \log(10)^{-1} \kappa) - n(< m_{\rm cut}) \nonumber \\
    &= 2 \left( 2.5 s - 1 \right) \kappa,
    \label{eqn:mag_cut}
\end{align}
where we have defined
\begin{equation}
    s =  \left( \frac{d\log_{10}n}{dm} \right)_{m_{\rm cut}}, \label{eq:log_slope}
\end{equation}
i.e. the magnification bias is dependent upon the logarithmic slope of the cumulative luminosity function. The factor of $1 - 2 \kappa$ in the first line reflects the geometric dilution of galaxies under magnification as they are spread out angularly. For more complicated surveys where the selection is no longer given by a simple magnitude cut the magnification bias is no longer simply given by the luminosity function slope $s$. However, we can still describe the effect more generally as
\begin{equation}
    \frac{\Delta n}{n} = 2 (\alpha - 1) \kappa, \label{eq:general_definition_alpha}
\end{equation}
to capture the linear response of the sample number density to foreground weak lensing. Here the minus one again captures the purely geometric effect whereby galaxies are angularly spread out by $\kappa$, while the coefficient $\alpha$ summarizes the total change in galaxies meeting the survey selection criterion when a magnification is applied (\cref{fig:mag_graphic}). For the case of a purely magnitude-limited survey, we would have $\alpha=2.5s$, but crucially there is no need to make this assumption as the magnification bias is purely a function of the survey selection for the galaxy sample, so in principle can be accurately determined from an accounting of this selection \citep{Krolewski2020,vonWietersheimKramsta2021,ElvinPoole2023,LRGSample}.
In \cref{sec:general_description} through \cref{sec:binwise} we describe how to estimate $\alpha$ while accounting for the full complexity of the selection for a realistic spectroscopic survey. 

\subsection{General description for realistic surveys}\label{sec:general_description}

The quantification of the magnification bias in realistic surveys requires a careful accounting of the linear effect of foreground magnification on the entire survey selection, rather than a single brightness limit, as in magnitude-limited surveys. The purpose of this section is to enumerate the physical effects contributing to the magnification bias in realistic surveys. Roughly speaking these can be categorized and computed in two groups: (a) the dropping out and falling in of galaxies \textit{across} the survey boundary and (b) the change in density \textit{within} survey boundaries, both due to magnification. Below, we will mathematically lay out the origin of these effects in spectroscopically confirmed samples of galaxies, with the purpose of extracting a robust estimator of the magnification bias; for related derivations and discussion of the case for imaging surveys we refer the reader to \citet{ElvinPoole2023}.

To discuss this generally, let the galaxies in the survey be described by variables $\lambda$, e.g. magnitudes, colours, shapes, etc. These variables all transform under the application of an external magnification, which we will denote by a lensing convergence $\kappa$. We thus need to properly account for the flow $d\vec{\lambda}/d\kappa$ of selection variables under changes in $\kappa$. Variables like the PSF and fiber magnitudes are not straightforwardly multiplied by a fixed factor but rather depend on the shape of the galaxy (see \cref{sec:lensing_mags}), and implicit selection criteria such as spectroscopic success rates (for redshift surveys) also need to be properly accounted for.

In practice, spectroscopic samples are targeted from an imaged \textit{target} sample with density $\rho(\lambda)$ in the $\lambda$ parameter space and success rate $r(\lambda)$, such that the number of spectroscopically confirmed galaxies in parameter volume $d\lambda$ and physical volume $dV$ is $n(\lambda) d\lambda dV = r \rho d\lambda dV$ and the number (density) within the selection criteria $\Omega$ is
\begin{equation}
    n_0 = \int_\Omega d\lambda\ r(\lambda) \rho(\lambda).
\end{equation}
This factorization is also useful for non-spectroscopic surveys to model catalogue completeness \citep{Krolewski2020}.

Under magnification, the galaxies shift by $\lambda \rightarrow \lambda + v(\lambda)$, where $v = (d\lambda/ d\kappa) \kappa$, such that the target distribution becomes at linear order 
\begin{equation}
    \rho_\kappa(\lambda) = \frac{\rho(\lambda - v)}{|\det J_V \det J_\lambda|} \approx (1 - 2\kappa) \rho - v \cdot \nabla_\lambda \rho - (\nabla_\lambda \cdot v) \rho, 
\end{equation}
where $J_\lambda$ is the Jacobian of this transformation and the factor of $|\det J_V|^{-1} = 1 - 2\kappa$ comes from the geometric dilution of the galaxies under lensing, i.e. the Jacobian factor for the physical volume element $dV$. The number density after magnification is then
\begin{align}
    n_\kappa &= (1 - 2\kappa) \int_\Omega r(\lambda) \rho_\kappa(\lambda)\ \nonumber\\
    &= (1 - 2\kappa) n_0 - \int_\Omega \left( \nabla_\lambda \cdot (r \rho v) - r \rho v \cdot \nabla_\lambda \ln r \right) \nonumber \\
    &= (1 - 2\kappa) n_0 - \int_{\partial \Omega} n v \cdot dT + \int_\Omega  n v \cdot \nabla_\lambda \ln r, \label{eq:lensed_number_density}
\end{align}
where we used the divergence theorem with $T$ being the boundary surface of the selection\footnote{Using non-standard notation for the boundary surface to avoid overuse of A and S.}. By plugging into the definition of $\alpha$ from \cref{eq:general_definition_alpha} we get
\begin{align}
\alpha = \frac{1}{2 n_0 \kappa} \left( - \int_{\partial \Omega} n v \cdot dT + \int_\Omega  n v \cdot \nabla_\lambda \ln r \right).
\label{eqn:surface_integrals}
\end{align}
We see that there are two distinct effects induced by the magnification on top of the geometric effect: the dropping out/in of galaxies from the selection boundaries as they are lensed, which we will denote as $\alpha_{s}$, and from any changes in the spectroscopic success rate within the selection volume itself, which we will denote as $\alpha_r$, with the net magnification bias $\alpha = \alpha_s + \alpha_r$. 

\subsection{Simple estimator for the magnification bias}
\label{sec:simple}

We now build a simple estimator for the magnification bias applicable to a galaxy catalogue, leveraging the general mathematical description in \cref{sec:general_description}. Any given catalogue of $N_0$ galaxies is a realization of the distribution $n(\lambda)$ such that we can replace the smooth integrals with sums over galaxies $i$ in the catalogues $V_{\rm obs}\int_{\Omega} n = \sum_i$.

The redshift success effect is an integral over the entire parameter volume and can be directly computed, independent of $\kappa$:
\begin{equation}
    \hat{\alpha}_r = \frac{V_{\rm obs}}{2 N_0 \kappa} \int_\Omega n v \cdot \nabla_\lambda \ln r = \frac{1}{2 N_0} \sum_{i=1}^{N_{g}} \frac{d\lambda}{d\kappa} \cdot \nabla_\lambda \ln r \label{eq:R}
\end{equation}
where we have replaced an integral weighted by $n$ with a sum over the galaxies in the spectroscopic catalogue. Given the functional form of the redshift success $r$, we can therefore estimate this contribution to the magnification bias summing over the galaxies directly. To constrain this one has to investigate the observations for a dependence of the spectroscopic success rate on any selection parameters. We discuss this for the BOSS samples specifically in \cref{sec:spectroscopic_success_rate} and will ultimately find that in this case, through the use of redshift failure weights, this term becomes negligible. 

The other contribution, the selection boundary effect, is a surface integral and can similarly be converted to a sum over galaxies in the catalogue. In order to do this we can first split the surface into patches where the dot product $v \cdot dT$ is positive and negative, which we will call $\partial \Omega_{\pm}$. The former region is where galaxies from within the selection boundary drop out, while the latter is where galaxies from outside drop in or, equivalently, drop out under the reverse ($-\kappa$) flow. Defining the numbers respectively as $N_{\pm}$ we can estimate
\begin{align}
    N_+ &= + V_{\rm obs} \int_{\partial \Omega_+} n v \cdot dT = N^s_0 - N^s_\kappa \nonumber \\
    N_- &= -V_{\rm obs} \int_{\partial \Omega_-} n v \cdot dT = N^s_0 - N^s_{-\kappa}, \label{eq:Nplusminus}
\end{align}
where we have defined $N^s_\kappa$ to be the number of galaxies in the spectroscopic samples that remain within its boundaries after applying a convergence $\kappa$. Therefore to estimate the surface integral, for a given sample, we can
\begin{enumerate}
    \item Transform $\lambda$ under magnification corresponding to $\kappa$.
    \item Reapply the selection and record how many drop out as $N_{+}$.
    \item Repeat for a transformation under $-\kappa$ to find $N_{-}$.
\end{enumerate}
Following this approach, the boundary effect is given by \citep{Krolewski2020,ElvinPoole2023}
\begin{equation}
    \hat{\alpha}_{s,\rm simple} = \frac{1}{2N_0 \kappa} \left(\hat{N}_- - \hat{N}_+  \right),
\end{equation}
where $N_\pm$ are computed using \cref{eq:Nplusminus}. We visualize the idea of this approach in \cref{fig:mag_graphic}. Since $N_\pm$ represent Poisson-sampled galaxies taken from distinct regions of width $\mathcal{O}(\kappa)$ near the survey boundary, the error on this estimate is, neglecting errors on the measured $\lambda$'s themselves, given by 
\begin{equation}
\sigma(\hat \alpha_{s,\rm simple})= \sqrt{\frac{N_- + N_+}{(N_0 2 \kappa)^2} + \frac{\hat{\alpha}_{s,\rm simple}^2 }{N_0}},
\end{equation}
where for completeness we also include the small additional Poisson noise of $N_0$, the size of the whole sample, in the second term.

It is important to note that this estimate incurs systematic errors of order $\kappa$ since we are effectively taking a simple one-sided derivative at the selection boundary, computing for $\pm \kappa$ only how many galaxies fall out or drop in, respectively. We are forced to do so because, starting from only a spectroscopically confirmed galaxy catalogue we inherently only sample the distribution $n(\lambda)$ within the selection boundary. If we also had access to galaxies from without we could evaluate the surface integral in \cref{eqn:surface_integrals} directly, i.e. simply by transforming all the targets under lensing by $+\kappa$ and recording the change $\delta N$ of galaxies within the selection boundary before and after \citep{Krolewski2020,ElvinPoole2023}
\begin{equation}
    \hat{\alpha}_{s,\rm simple}^{\rm target} = \frac{\delta N}{2N_0 \kappa}.
\end{equation}
In addition, this procedure can be performed also in the $-\kappa$ direction, allowing us to take a central difference that leads to more suppressed errors of order $\kappa^2$ \citep{Samuroff2019}.

Combining the contributions from the photometric selection and the spectroscopic success rate our simple estimator for the magnification bias is therefore
\begin{equation}
    \hat{\alpha}_{\rm simple} = \frac{N_\kappa - N_0}{2 N_0 \kappa} = \hat{\alpha}_{s,\rm simple}  + \hat{\alpha}_r.
    \label{eq:mag_bias_estimate_simple}
\end{equation}
\cref{eq:mag_bias_estimate_simple} is the main conceptual tool we will use in this paper and is the equivalent of Eq. (24) in \citet{ElvinPoole2023} for spectroscopic surveys. However, in order to proceed further we will need to further refine our estimator to marginalize over higher-order dependence on the stepsize $\kappa$ in $\hat{\alpha}_{s, \rm simple}$.

\subsection{Binwise estimator for $\hat{\alpha}_s$}
\label{sec:binwise}

Naively, the simple estimate \cref{eq:mag_bias_estimate_simple} becomes arbitrarily precise as the step size $\kappa$ becomes large due to the ever-increasing number of galaxies dropping in and out of the boundaries. However, the simple estimate of $\alpha_s$ constitutes a numerical one-sided derivative with linear error as a function of step size $\kappa$ that should be properly accounted for. In particular, since this linear correction is of unknown size, it is appropriate to marginalize over its contributions as we increase the stepsize. In this subsection, we introduce a simple binwise method to this effect.

One possible method to account for the linear effect is to construct a series of $\hat{\alpha}_s$ measurements of varying step size $\kappa_i$. However, due to the overlap in dropped-out galaxies, these estimates will tend to be highly correlated. As a simple alternative, we can choose to estimate the function in bins which gives us independent samples. Using a series $\kappa_i$ of step sizes with $\kappa_0 = 0$ we can get an estimate in each bin as
\begin{equation}
    A_i \equiv \frac{(N_{\kappa_{i}} - N_{\kappa_{i-1}})- (N_{-\kappa_{i}} - N_{-\kappa_{i-1}}) }{2 N_0 (\kappa_i-\kappa_{i-1})}. \label{eq:mag_bias_A_definition}
\end{equation}
Each measurement then represents an independent sample for the bin $\kappa_{i-1}$ to $\kappa_i$ with uncertainty
\begin{equation}
    \sigma_{A_i}^2 = \frac{|N_{\kappa_{i}} - N_{\kappa_{i-1}}| + |N_{-\kappa_{i}} - N_{-\kappa_{i-1}}|}{(2 N_0  (\kappa_i-\kappa_{i-1}))^2}.
\end{equation}
We can reasonably neglect the correlated uncertainty in $N_0$\footnote{The error on the measurement from the uncertainty of the sample size $N_0$ is given by $\alpha/\sqrt{N_0}$ and is approximately 0.003 for the CMASS sample and 0.004 for the LOWZ sample. Added in quadrature to the statistical uncertainty of our measurements presented in \cref{tab:baseline_result}, the additional term would affect the error by less than 2\%. }.
To find the magnification bias we fit a linear function to account for the systematic error to first order
\begin{equation}
    A_i = \hat \alpha_s - \beta_{\rm sys} \frac{\kappa_{i-1}+\kappa_i}{2}, \label{mag_bias_estimate}
\end{equation}
where $\hat \alpha_s $ is our estimate of the magnification bias in the sample and $\beta_{\rm sys}$ is the slope of the systematic error. The error for $\hat \alpha_s $ is then the uncertainty on the fit parameter. In this formalism, the range of $\kappa$ step sizes should be chosen based on the goodness of fit for the linear curve, such that further beyond-linear variations in $n(\lambda)$ can be neglected. If a two-sided derivative can be taken as described above the systematic term has to be modified to be quadratic and should be less significant. We note that this approach also highlights numerical issues near the photometric boundary which would show up as offsets for the smallest $\kappa_i$ bins from the rest of the fit. Again the estimate can be combined with the contribution from the spectroscopic success rate to obtain the full magnification bias of the sample $\hat \alpha = \hat \alpha_s + \hat \alpha_r$.

\subsection{Impact on observables}

After having described in detail how to estimate the change in \textit{observed} local number density due to the magnification bias $\alpha$, let us briefly review the contributions of the magnification effect to cosmological observables.
The change in \textit{observed} local number density results in an additional contribution to the \textit{observed} clustering of galaxies due to lensing by matter in the foreground of each galaxy. For the angular 2-point function between galaxies and some other observable $B$, we can write generically using the Limber approximation \citep{Limber},
\begin{align}
    \hat C^{gB}_\ell& = C^{gB}_\ell + \int \frac{d\chi}{\chi^2} W^\mu(\chi) W^B(\chi) P_{mB}\left(k=\frac{\ell+\frac12}{\chi},z\right),
    \label{eq:cells}
\end{align}
where $P_{mB}$ is the power spectrum between the observable $B$ and the nonlinear matter density, for wavenumber $k$ and multipole $\ell$ at comoving distance $\chi$ corresponding to redshift $z$. $W^B$ the kernel of observable $B$ and $W^{\mu}$ is the magnification kernel. We neglect corrections to the Limber approximation on large scales for simplicity of presentation \citep{Loverde08}.

The magnification kernel is given by the combined lensing kernels $W^\kappa(z,z')$ of galaxies in the sample, which act as background galaxies that get lensed along the line of sight, and scaled by the change in the observed number density 
\begin{align}
    W^\mu(\chi) = \int_{z(\chi)}^{\infty} dz'\,
      2 \,\left(\alpha(z') - 1 \right)\ \frac{dN(z')}{dz'}\ W^\kappa(z(\chi),z'). 
    \label{eq:zdep_wmu}
\end{align}
In many analyses, redshift evolution for the magnification bias within a galaxy is ignored such that $\alpha$ can be pulled out of the integral. We will discuss this assumption in detail for SDSS BOSS in \cref{sec:redshift_evolution}. 
The lensing kernel for a source at $z'$ is given by 
\begin{equation}
    W^{\kappa}\left(z, z'\right)=\frac{3}{2} H^2_{0} \Omega_{\rm m,0} (1+z)
    \chi(z)\left(1- \frac{\chi(z) }{ \chi ( z')}\right).
\end{equation}
For cross-correlations with CMB lensing the kernel for $B$ is given by $W^{\kappa}(z, z_{\rm CMB})$ where $z_{\rm CMB}$ is the source redshift of the CMB. 

For the galaxy auto-correlation, the effect of magnification bias is overall smaller. Under the limber approximation, the change is given by
\begin{align}
    \hat C^{gg}_\ell &= C^{gg}_\ell + 2 \int \frac{d\chi}{\chi^2} W^\mu(\chi) W^g(\chi) P_{mm}\left(k=\frac{\ell+\frac12}{\chi},z\right) \nonumber \\ 
    &+ \int \frac{d\chi}{\chi^2} W^\mu(\chi) W^\mu(\chi) P_{mm}\left(k=\frac{\ell+\frac12}{\chi},z\right),
    \label{eq:cells_gg}
\end{align}
where $W^g(\chi)= \frac{dN(z)}{dz} \frac{dz}{d\chi}$.

The magnification bias terms in \cref{eq:cells,eq:cells_gg} amount to a nontrivial foreground contribution to observed galaxy angular cross-correlations that must be properly modelled in addition to the galaxy-galaxy and galaxy-matter cross-correlations at the positions of the galaxies themselves probed by these statistics. We investigate the size of the effects for SDSS BOSS in \cref{sec:result_baseline}.

\section{Lensing of the measured magnitudes}
\label{sec:lensing_mags}

In the previous section, we derived, under quite general assumptions, how foreground magnifications alter the selection of galaxies in realistic surveys. An important ingredient in this derivation is the change of observed galaxy properties under lensing. We characterized this generally as the derivative $d\lambda/d\kappa$, describing the flow of (observed) galaxy properties under nonzero magnification. 
For SDSS BOSS the observables used in the photometric selection are the cModel, model, fiber2, and PSF magnitudes. In this section, we show how these observed magnitudes change under lensing to first order in $\kappa$. These derivations apply to equivalent magnitudes in other surveys, but any additional observables used for photometric criteria in those surveys would also need to be considered. It is convenient to work with the observed fluxes and then convert the magnified fluxes to magnitudes.

We can describe galaxies with an intensity profile $I_{\theta_e}(\theta, \phi)$ with a characteristic angular radius $\theta_e$ so that the profile only depends on $\theta/\theta_e$, where $\theta$ is the radial angular coordinate from the profile centre and $\phi$ the azimuthal angle. Under lensing the surface brightness is constant but the shape changes as $\theta_e \rightarrow \theta_e (1+\kappa)$. We can write this as 
\begin{equation}
    I_{\theta_e} \rightarrow I_{\theta_e}\left[ \frac{\theta}{(1+\kappa)}, \phi \right].
\end{equation}
The primary cModelMag and modelMag magnitudes used in \citet{Reid2016} capture the full light observed from the galaxy: $F = \iint I_{\theta_e}(\theta, \phi) \theta d\theta d\phi$. Plugging in the lensed profile and using a variable change one can show that to first order in $\kappa$ the fluxes get lensed as $F_\kappa = (1 + 2\kappa ) F$, i.e. as expected by the magnification $\mu$.

For the $\ang{;;2}$ fiber magnitudes (fib2Mag) and PSF magnitudes (psfMag), we additionally need to account for the shape change of the galaxy under lensing. The result will now depend on the characteristic angular radius $\theta_e$ of the galaxy. We assume a radially symmetric de Vaucouleurs profile for the Intensity $I_{\theta_e}(\theta)$ as a function of angular distance from the galaxy centre with de Vaucouleurs radius $\theta_e$ given by the SDSS survey. The lensing of the profile is as before but we need to carefully consider how the fluxes are measured. For the fiber flux, there is a cutoff at an angular distance of $ \bar \theta = \ang{;;2}$. For the PSF flux a Gaussian profile with $\sigma_{\rm PSF}$, converted from the PSF\_FWHM values provided by the survey (see \cref{sec:SDSS_data}), is used. We have for the fiber flux at cutoff $\bar \theta$ and the PSF flux 
\begin{align}
    F^{\textrm{fib} \bar \theta} &= 2\pi \int_0^{\bar\theta} I_{\theta_e}(\theta) \theta d\theta ,\\
    F^{\rm PSF} &= \frac{1}{C} \int e^{- \theta^2/(2\sigma_{\rm PSF}^2)} I_{\theta_e}(\theta) \theta d\theta ,
\end{align}
where $C$ is a normalization constant.
After some algebra, the change in observed flux for the two fluxes is, to first order in $\kappa$,
\begin{align}
    \frac{F_\kappa^{\rm fib2}}{F^{\rm fib2}} &=  1 + \left(2- \left. \frac{\dd \ln F^{\rm fib \bar \theta}}{\dd \ln \bar \theta}\right|_{\bar \theta=\ang{;;2}} \right) \kappa  , \label{eq:fiber2_mag_change}\\
    \frac{F_\kappa^{\rm PSF}}{ F^{\rm PSF}} &= 1 + \left(2- \left. \frac{\dd \ln F^{\rm PSF}}{\dd \ln \sigma}\right|_{\sigma=\sigma_{\rm PSF}} \right) \kappa . \label{eq:psf_mag_change}
\end{align}
Much like the magnification bias itself the magnification of these size-dependent magnitudes gets a correction for the purely geometric factor of $2$ by the effective slope of the light profile at the sampled angular scale.
The term in the brackets, $\left( 2 - \frac{d \ln F}{d \ln X} \right)$,  we define as the $\kappa$ multiplier.

\begin{figure}
\includegraphics[width=\columnwidth]{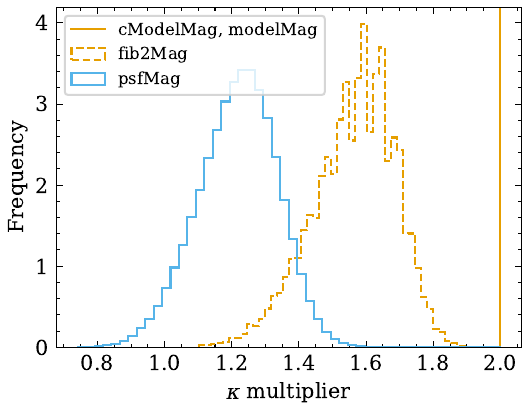}
\caption{The distribution of the $\kappa$ multipliers for different observed magnitudes for the CMASS sample in the $i$ band. For modelMag and cModelMag the lensing increases the corresponding flux by a factor (1+2$\kappa$), a $\kappa$ multiplier of exactly 2 (shown as a solid vertical orange line). For the psfMag and fib2Mag, we need to apply corrections that account for the size change in the galaxy, significantly reducing the increase in flux under lensing from a $\kappa$ multiplier of 2 to a lower value depending on the galaxy's shape and the magnitude considered as derived in \cref{sec:lensing_mags}. We show the normalized histograms of $\kappa$ multipliers for galaxies in the CMASS sample as a blue line for the psfMag and an orange dashed line for the fiber2Mag.}
\label{fig:lensing_multipliers}
\end{figure}

\cref{fig:lensing_multipliers} shows the distribution of these $\kappa$ multipliers for the different magnitudes in CMASS. We calculate the corrections numerically for each galaxy. When the full light of the galaxy gets captured the change of observed flux scales as $2\kappa$. However, for the fiber and PSF flux, this multiplier 2 gets reduced by a log derivative as derived in \cref{eq:fiber2_mag_change,eq:psf_mag_change}. Characteristic values for the multiplier onto $\kappa$ are around 1.5 for the \ang{;;2} fiber flux and 1.2 for the PSF flux in our case. We note that they are similar for the other BOSS samples. Since the effective slope effectively reduces the magnification effect for the fiber and PSF magnitudes we find that these corrections are crucial to not overestimate the impact of the photometric selections including the $\ang{;;2}$ fiber magnitude and PSF magnitude. We discuss this in \cref{sec:results_with_simple_estiamtor}. 

Throughout this work, we will make the fiducial assumption that galaxies follow a deVaucolouleurs profile. Assuming a deVaucouleurs profile for our galaxies is a modelling choice that can introduce systematic bias to the estimate if galaxies deviate from it significantly. To quantify the impact we also repeat the analysis assuming an exponential profile using the same half-light radius, assuming that the half-light radius is a better-measured quantity than the particular profile shape. We will use the difference as an estimate for the systematic uncertainty in our estimate. An alternative method, proposed in \citet{LRGSample}, is to perform the light-profile derivatives above directly at the level of the data by measuring the distribution of fluxes relative to the shape parameters for a given type of galaxy in the particular survey. This method foregoes the exact analytic expressions for the impact of magnification on galaxy light profiles described above but instead uses the statistical trends in the data itself to determine the magnification effect, which could serve as a way to calibrate the systematic error from light-profile modelling. In addition, while we will work with radial profiles throughout this work, since the BOSS LSS catalogue does not provide us with further shape information, in principle our analysis could be extended to include the effect of galaxy ellipticities as well; we save this exploration for future work.

\section{Application to SDSS Data}\label{sec:application_SDSS_data}

\begin{figure}
\includegraphics[width=\columnwidth]{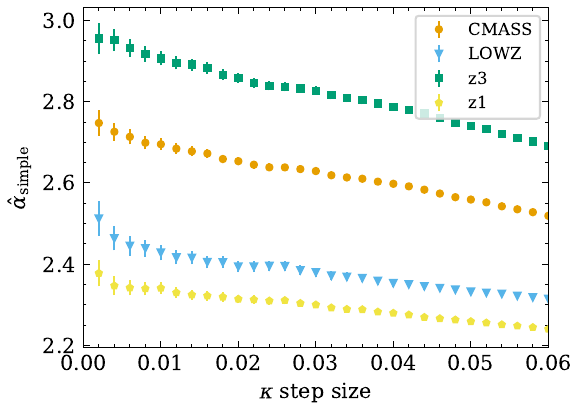}
\caption{Results of the simple estimate of the magnification bias $\hat \alpha_{\rm simple}$ for a range of step sizes in $\kappa$. We show the results for the CMASS sample as orange dots, for LOWZ as blue triangles, z3 as green squares and z1 as yellow pentagons. For larger step sizes the statistical constraining power improves however the systematic error of the estimate increases. The measurements for different step sizes are correlated, to get the most accurate measurement of the magnification bias while accounting for the systematic error we leverage a binwise estimator instead. \label{fig:alpha_simple}}
\end{figure}

In this section, we present the results for the magnification bias measurements for the SDSS BOSS survey. In \cref{sec:results_with_simple_estiamtor} we first present the results with the simple estimator which allows us to discuss the tradeoff between statistical and systematic uncertainty and the impact of individual criteria in the photometric selection. We consider the effect of spectroscopic success rates for BOSS in \cref{sec:spectroscopic_success_rate}. In \cref{sec:result_baseline}, we describe our baseline results for the magnification bias in the BOSS samples with our final estimator and show the impact on cosmological observables. We also perform a detailed accounting of the statistical and systematic uncertainty as well as the dependence on various analysis choices. We investigate two of these in detail: the evolution of the magnification bias with redshift in \cref{sec:redshift_evolution} and the impact of common cosmological analysis choices in \cref{sec:impact_analysis_choices}.

\subsection{Results with the simple estimator} \label{sec:results_with_simple_estiamtor}

We first consider the results of a simple estimator $\hat \alpha_{\rm simple}$ as defined in \cref{eq:mag_bias_estimate_simple}. 
This allows us to get an estimate of the magnification bias for one fixed lensing $\kappa$ step size applied to the data. As we will argue in \cref{sec:spectroscopic_success_rate} the spectroscopic success rate contribution is negligible with the BOSS samples, as defined, so we will focus on the selection boundary effect below. The goal is to show and discuss the tradeoff between statistical constraining power and systematic bias inherent to using a fixed step size. The simple estimator will also allow us to discuss how different photometric criteria impact the result.

As described in \cref{sec:simple} the stepsize $\kappa$ is the only free parameter in the simple estimator---taking a large step size causes more galaxies to drop out of the selection boundary post-lensing, reducing the Poisson errors but also increasing the nonlinear corrections to the finite difference. In \cref{fig:alpha_simple} we show the magnification bias $\hat \alpha_{\rm simple}$ as a function of $\kappa$ step sizes. Towards larger step sizes, the Poisson errors decrease but, perhaps more significantly, the estimates show a systematic trend towards larger stepsizes, such that even at $\kappa = 0.01$ the simple estimate is more than $1\sigma$ different than the estimate using $\kappa = 0.002$. We note that the difference is even larger than it naively appears due to the correlation between the stepsizes. The dominant underlying reason for the systematic offset, as pointed out in our derivation in \cref{sec:simple},  is that this simple estimator constitutes a one-sided derivative at the boundary of the photometric selection. This derivative will incur an error of order $\kappa$ which, given our small statistical uncertainty, can not be neglected and instead needs to be accounted for. This motivates adopting the binwise estimator presented in \cref{sec:result_baseline} where such trends can be effectively marginalized over. We highlight that previous analyses typically used a single fixed step size, not marginalizing over this systematic bias \citep[see e.g.][]{Singh2019,Krolewski2020}. Especially for large step sizes the statistical error appears to decrease while the systematic bias increases which can lead to a systematic bias of multiple $\sigma$   in the estimate. 

\begin{table}
\include{Tables/photometric_criteria}
\caption{Overview of the impact on our estimate of the magnification bias for each photometric selection criteria for CMASS and LOWZ for a fixed step size of $\kappa =0.01$. The values do not necessarily sum to our estimate of $\hat \alpha_{\rm simple}$ since multiple criteria can affect the same galaxy. The variables used for the photometric selection are introduced in \cref{sec:SDSS_data} and \cref{fig:color_selection_CMASS} visualizes part of the selection. \label{table:magnification_bias}}
\end{table}

In order to understand the role of the selection function on the magnification bias it is instructive to consider the impact of the various photometric selection criteria on $\alpha$ individually. This is more intuitive when considering $\hat{\alpha}_{\rm simple}$ since we can simply record how many galaxies fail any given photometric criteria after applying the magnification. The impact of each selection criteria is presented in \cref{table:magnification_bias}. The table clearly demonstrates that the overall CMASS and LOWZ samples can not be taken as magnitude-limited. Assuming they are would significantly underestimate the total magnification bias in the sample. For our case, we would only find around $\alpha \approx 0.9$ for CMASS and $\alpha \approx 0.3$ for LOWZ. The CMASS and LOWZ samples are predominantly limited by a diagonal cut in $i_{\mathrm{cmod}} - d_{\perp}$ and $ r_{\mathrm{cmod} } - c_{\parallel}$ space respectively (see \cref{fig:color_selection_CMASS}). These diagonal cuts contribute the majority of the overall magnification bias and need to be included for an accurate estimate.
Since the z1 and z3 samples are a direct reselection of the same data they can also not be taken as magnitude limited. 

\Cref{table:magnification_bias} also shows that the cut on the fiber2 magnitude ($i_{\rm fib2}$) for CMASS, to ensure a reasonable spectroscopic success rate, has a non-negligible impact of around 0.2 on the overall magnification bias. This shows that one has to consider the full set of criteria employed in the selection function to accurately predict the magnification bias effect. For the brighter LOWZ galaxies, no such fiber2 magnitude cut was applied. 

The magnification bias due to cuts using the PSF magnitudes ($i_{\rm psf}, z_{\rm psf}$) is smaller overall though they do contribute above the level of our statistical uncertainty for CMASS. The underlying reason for this is that under lensing and de-lensing the quantities $i_{\rm psf}-i_{\rm mod}$ and $z_{\rm psf} - z_{\rm mod}$ move predominantly along the selection boundary rather than across it. This is why few objects drop out and therefore the impact on the overall magnification is small. 

For the magnification effect of the fiber2 and PSF magnitudes, it is important to correctly account for the shape change under lensing (\cref{sec:lensing_mags}) in order to correctly model the flow in selection space under lensing. Not correcting for the shape change would overestimate the lensing magnification of these magnitudes. We investigate the impact of this specifically for the CMASS sample and our simple estimator with fixed step size: when completely neglecting the shape change of galaxies the impact of the cut on the fiber2 magnitude shifts from +0.26 to +0.38. For the cuts involving the PSF magnitudes the one in the i-band only slightly shifts from +0.01 to +0.02, however in the z-band the impact shifts from -0.05 to -0.14. 
For other surveys, this does not have to be similar since it sensitively depends on the functional form of the applied photometric selection.

\subsection{Impact of the spectroscopic success rate}\label{sec:spectroscopic_success_rate}

While the photometric selection dominates the overall magnification bias there can be an additional impact from the spectroscopic observations if the success rate of spectroscopic observations depends on the observed brightness which is affected by lensing \cref{eq:R}. This requires careful consideration for a specific survey and it is well known that the redshift success rate of the BOSS samples is well-described as a function of the fiber2 magnitude as shown in Fig. 7 in \citet{Reid2016}. 

Based on this one could estimate the impact on the magnification bias for the raw LOWZ and CMASS samples as in \cref{eq:R}. The success rates for CMASS and LOWZ are well-fitted by a sigmoid function. We would find $\hat \alpha_r \approx 0.05 (\hat \alpha_r \approx 0.00)$, together with different $\hat \alpha_s$, for CMASS (LOWZ) for the catalogues without any weights. 
However, for cosmological analyses, SDSS employs weights that include a correction for this effect. Specifically, a redshift failure weight is used. To retain the clustering impact of galaxies for which no redshift could be obtained the nearest neighbor is up-weighted. This alone would not account for the fiber2 magnitude dependence of the failure rate since mostly brighter galaxies will be up-weighted. However, an additional step is taken where the weights get modified to match the distribution in fiber2 magnitude of the failed galaxies. 
This corrects for the dependence of the spectroscopic success rate on the fiber2 magnitude by creating an effective sample where the spectroscopic success rate does not depend on the selection parameters. Therefore applying the recommended weights to the data should make the contribution of $\hat \alpha_r$ negligible for the BOSS LSS catalogs. Subsequently, we will consider the photometric magnification bias $\hat \alpha_s$ and the full magnification bias $\hat \alpha$ as equivalent and neglect $\hat \alpha_r$.

\subsection{Baseline results for magnification bias} \label{sec:result_baseline}

\begin{figure}
\includegraphics[width=0.95\columnwidth]{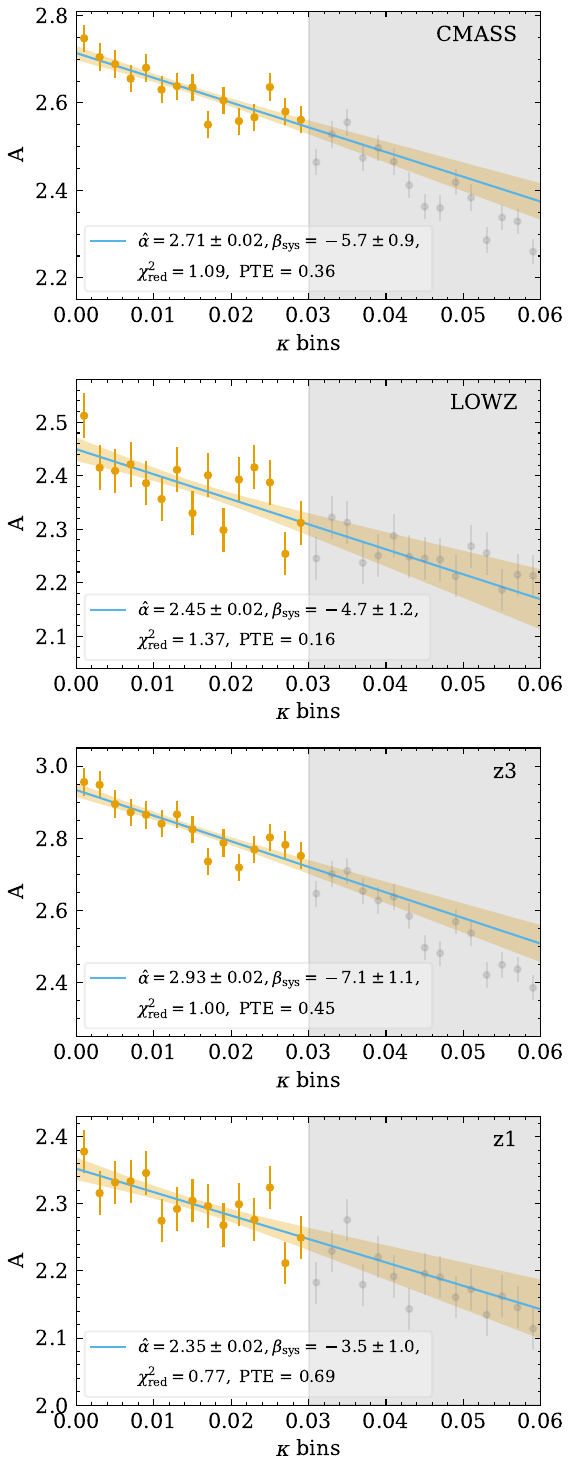}
\caption{Main result of the work showing our estimates for $\hat \alpha$ using our binwise estimator. Shown is $A$ as a function of bins in lensing convergence, $\kappa$ as dots with individual measurement errors, for each SDSS sample: [from top to bottom] CMASS, LOWZ, z3, and z1. We fit a line up to $\kappa=0.03$ to estimate $\hat \alpha$ and the slope of systematic error $ \beta_{\rm sys}$. We show the best fit as a solid blue line and show a one-sigma uncertainty band in shaded orange. For each fit, we list the fitted values as well as the reduced $\chi^2$ and PTE value. Datapoints used for the fit are in orange and additional points not used in the fit are shown in grey to indicate that the fit is well within the linear regime. \label{fig:alpha_estimate}}
\end{figure}

We now present results for the magnification bias for the four main BOSS LSS samples using the binwise estimator presented in \cref{sec:binwise}. The fits from which our estimates are derived are shown in \cref{fig:alpha_estimate}, and our results are summarized in \cref{tab:baseline_result}. We find good PTE values (PTE > 0.05) for a linear fit with $\kappa_{\rm max} \in [0.01, 0.02, 0.04, 0.05]$ for all 4 samples. In this range, the results are statistically consistent with each other. For larger $\kappa_{\rm max}$ especially CMASS and z3 are not well described by a linear fit, indicating that at this scale the linear approximation becomes unreasonable. We use a conservative cutoff of $\kappa_{\rm max} = 0.03$ to remain well within the linear regime.

\begin{table}
\include{Tables/baseline_results}
\caption{Baseline result for the magnification bias using the binwise estimator with statistical and systematic uncertainties. Listed are the best estimate value, the statistical uncertainty of the fit, and the systematic uncertainty from our choice of light profile. Additionally one also has to consider the impact of redshift evolution within the sample which we discuss in \cref{sec:redshift_evolution}. The results are sensitive to the exact analysis choices, \cref{table:analysis_chocies} provides results when using other commonly used choices. \label{tab:baseline_result}}
\end{table}

\begin{figure}
\includegraphics[width=\columnwidth]{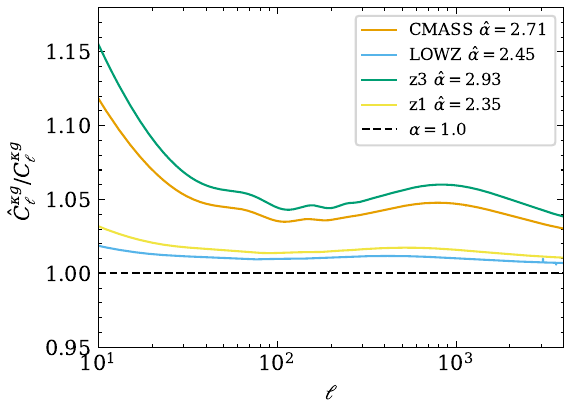}
\includegraphics[width=\columnwidth]
{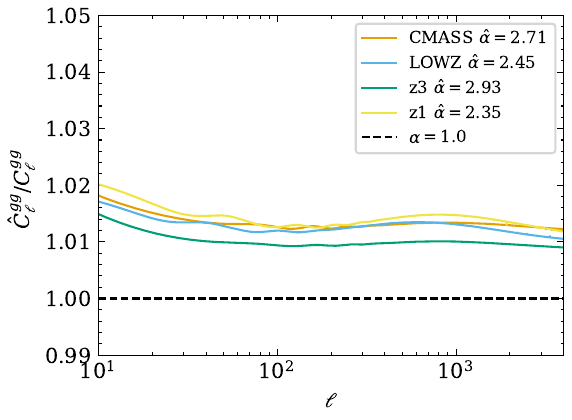}
\caption{The fractional shift of the expected galaxy-CMB lensing cross-correlation ($\kappa g$) [Upper] and galaxy auto-correlation  (gg) [Lower] when the magnification bias is included ($\hat{C}_\ell$) relative to when it is excluded ($C_\ell$) for four SDSS samples: CMASS [orange], LOWZ [light blue], z3 [green], z1 [yellow]. Used are the best-fit redshift-averaged values for $\alpha$ for each sample. The prediction with no contribution from magnification bias ($\alpha=1$), close to what one would obtain if using a simple magnitude-cut estimate, is also shown [black dashed].}
\label{fig:impact_autocross}
\end{figure}

Based on our best-fit values for the magnification bias we can quantify the impact on cosmological observables resulting from this effect. \cref{fig:impact_autocross} shows the fractional contribution of the magnification bias to the angular two-point auto-correlation of BOSS galaxies and galaxy-CMB lensing cross-correlation\footnote{Using \cref{eq:cells,eq:cells_gg}. For these calculations, we use the \ccl code \citep{Chisari2019} and assume a fiducial cosmology matching the CMB+BAO constraints from \cite{PlanckCollaboration2018} and a galaxy bias of 2. We are using pyccl after version 2.5.1 for which a critical bug in the magnification bias module was corrected.}. We find that the effect of the magnification bias is quite significant for the CMB lensing-galaxy cross-correlation. For the CMASS and z3 samples, the effect leads to a $\sim$5\% enhancement at $\sim\ell>100$ and up to $15\%$ at large angular scales. The effect is smaller for the LOWZ and z1 samples, around 2-3\%, due to there being less intervening matter at low redshifts. The impact of the magnification biases on the galaxy auto-correlation is smaller around 1-1.5\% and comparable across the four samples. While the effect is smaller for the auto-correlation it will become increasingly statistically relevant for the improved constraining power in upcoming surveys. Significantly, using the simple magnitude cut measurement \cref{eqn:mag_cut} would incorrectly predict $\alpha \lesssim 1$ for the high-redshift samples, leading to a small suppression in the signals and resulting in a $5-10\%$ under-estimation of the CMB lensing cross-correlation.

Our statistical uncertainty---on the order of $1\%$--- in the baseline results is small. The overall uncertainty is instead dominated by systematic errors and the exact value of the magnification bias depends sensitively on a range of analysis choices. To quantify these we investigate a range of sources for systematic error and modeling choices to give a realistic accounting.

While most of our estimation is Poisson limited, there is a caveat. For the estimate of the change in observed fiber and PSF magnitudes, we needed to assume a light profile. The modelling choice of light profile is an inherent uncertainty to our method.
To estimate the systematic uncertainty of our modelling choice of using de Vaucouleurs profiles, we recalculate the estimate with exponential profiles instead. We then use the difference between the two estimates as our systematic error budget. For the LOWZ sample, this modelling choice has little effect because the photometric selection cuts using PSF Magnitudes have little effect overall on the magnification bias estimate. For CMASS the PSF and fiber2 magnitude cuts shift the inferred magnification bias in opposite directions leading to a partial cancellation. Due to the different redshift cuts, this cancellation does not happen for z1 and z3 and we find a larger effect in these samples (column ``Light profile choice'' in \cref{table:analysis_chocies}). 

The estimate presented measures the redshift-averaged magnification bias for the sample. While the measurement of the average is precise, this can introduce a systematic error for a cosmological inference if there is a significant evolution in redshift. The reason for this is that magnification bias at higher redshift has a larger effect on the observed correlation functions since there is more intervening matter along the line of sight. We find significant dependence on redshift for CMASS which we discuss in detail in \cref{sec:redshift_evolution} and we discuss options to account for this redshift evolution in cosmological analyses.

Finally, the results are sensitive to the exact analysis choices made for a given cosmological inference. In \cref{sec:impact_analysis_choices} we provide a range of results for common alternative analysis choices used in the literature. For a specific cosmological analysis, it is advisable to match one of these cases exactly or recalculate the estimate for the specific set of analysis choices used in the analysis. Most of the differences we find can be attributed to the redshift evolution within the samples which we discuss next.

\subsection{Redshift evolution}\label{sec:redshift_evolution}

\begin{figure}
\includegraphics[width=\columnwidth]{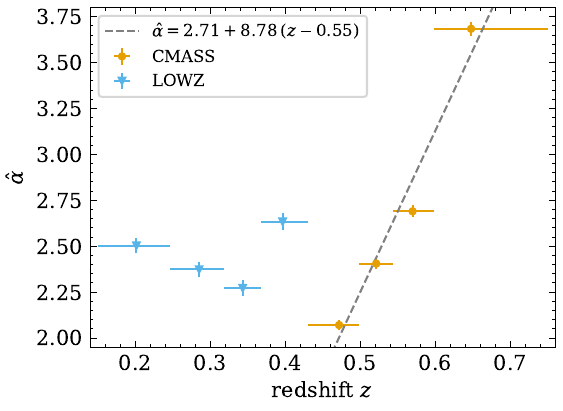}
\caption{Redshift dependence of the magnification bias in CMASS and LOWZ, each split into 4 equal-weight subsamples in redshift. The measurement for each redshift bin is shown as orange dots for CMASS and blue triangles for LOWZ, with the lines indicating the redshift range of the subsample and the uncertainty of the measurement. The clear trend with redshift in CMASS is shown by the dashed line. }
\label{fig:redshift_evolution}
\end{figure}

\begin{figure}
\includegraphics[width=\columnwidth]{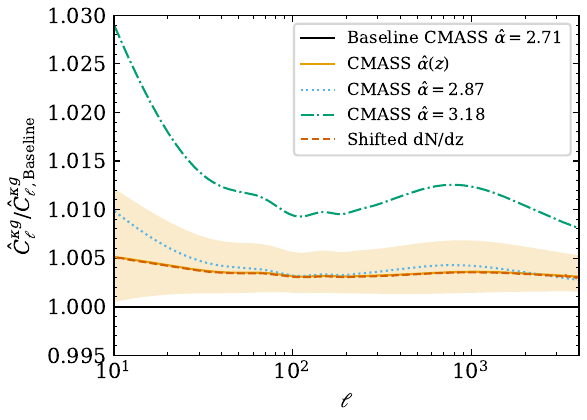}
\includegraphics[width=\columnwidth]{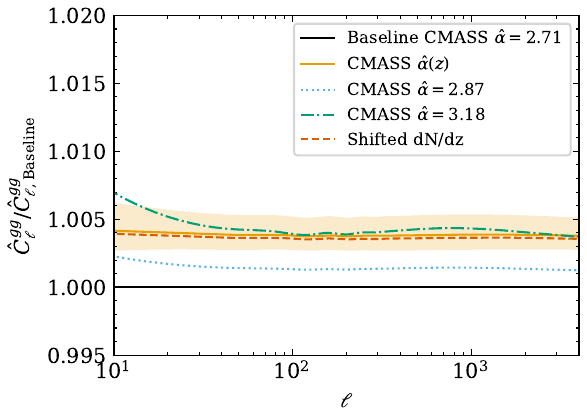}
\caption{Observables when accounting for the redshift dependence of magnification bias in CMASS compared to using the averaged magnification bias value for the cross-correlation with CMB lensing [upper panel] and auto-correlation [lower panel] is shown as an orange line in each. We show a shaded orange band for the uncertainty in the linear evolution. We also show the relative change for 3 constant magnification bias values: our baseline value of $\hat \alpha = 2.71$  which gives a ratio of 1 shown as a black line as well as $\hat \alpha = 2.87$ and $\hat \alpha = 3.18$ shown as blue dotted and green dash-dotted lines respectively. Finally, we also show the relative change when applying a shift in $dN/dz$ as a red dashed line which closely matches the result for the redshift-dependent magnification bias. Further details can be found in \cref{sec:redshift_evolution}.}
\label{fig:cross_corr_redshift_evolution}
\end{figure}

Up to this point, we have described the magnification bias for a given galaxy sample with a redshift-independent constant $\alpha$. However, the magnification effect is due to the cumulative lensing of a galaxy by all foreground matter, leading to two separate sources of redshift dependence in the magnification bias kernel: (a) higher-redshift galaxies in a given sample are weighted more due to there being more foreground lensing and (b) redshift evolution within the sample can make lead to a redshift-dependent $\alpha(z)$. The former effect is automatically taken care of by integrating over all galaxies in the sample for the magnification kernel, but the latter is typically ignored and a constant $\alpha$ factored out of the redshift integral in \cref{eq:cells}. This is a zeroth-order approximation expanding in the redshift dependence of the sample, and higher-order effects in the redshift derivatives $\alpha^{(n)}(z)$ will generally produce both a change in amplitude and scale dependence.

Spectroscopic surveys like BOSS present an opportunity to more fully model this effect. Since our estimation is performed on spectroscopic galaxies, we can simply further split the observed galaxies by redshift to measure $\alpha(z)$. In \cref{fig:redshift_evolution} we show such a measurement for LOWZ and CMASS, each split into four redshift bins with equal weighted numbers of galaxies. LOWZ exhibits only small differences between different bins and no clear trend. CMASS galaxies show a strong, monotonic trend in redshift, with $\alpha$ increasing by almost a factor of two between our lowest and highest redshift bins. This trend is well-described by a linear fit to the data. For readability, we centre the line at the weighted mean redshift of the whole sample ($z = 0.55$) where we recover our fiducial mean measurement, giving us
\begin{equation}
    \alpha_{\rm CMASS}(z) = 2.71 + 8.78 (z - 0.55),
    \label{eq:cmass_zdep}
\end{equation}
which can be input into the calculation of observables\footnote{In \cref{fig:cross_corr_redshift_evolution,fig:magnification_kernels} we use one additional significant figure of precision of our fit $\hat \alpha(z) = 2.708 + 8.775 (z-0.552)$ to visually better show how well it matches with the shift in $dN/dz$. Due to the overall uncertainty, we round the results listed in \cref{eq:cmass_zdep}}. The scatter of the points implies uncertainties for the constant and slope of 0.08 and 1.26. Note that here we focus on the overall sample, if narrow redshift bins of CMASS are analyzed instead, non-linear evolution may become relevant, and finer binning should be investigated. For example \citet{Krolewski2020}
report that at the high redshift tails ($z>0.65$) the increase is more than linear.

\cref{fig:cross_corr_redshift_evolution} shows the change in $\hat C_\ell^{\kappa g}$ when including the linear redshift evolution we measure for CMASS \cref{eq:cmass_zdep} compared to using a constant redshift-averaged magnification bias. For the cross-correlation with CMB lensing the change is sub-percent and small compared to the overall 5-10\% change due to magnification bias. We highlight that this is not a systematic bias of our redshift-averaged $\hat \alpha$ measurement, but rather a physical consequence of the redshift-dependent impact of the magnification bias on observables.

Considering the fiducial cross-correlation with CMB lensing alone, it appears the shift is reasonably well described by an increase in the constant magnification bias. When approximately translate the freedom in the fit for the linear evolution (orange band in \cref{fig:cross_corr_redshift_evolution}) into a range for the correction we find a correction for $\hat \alpha_{\rm CMASS}$ of $+0.16 \pm 0.10$ due to redshift evolution for cross-correlations with CMB lensing. However, we caution that the same shift is not applicable for the auto-correlation (as shown in \cref{fig:cross_corr_redshift_evolution}), which requires a significantly larger shift in $\alpha$, and in addition will also depend on both the cosmology and model of galaxy clustering used to model the correlation functions. Therefore modeling the magnification bias as a constant for a combined analysis, e.g. a 3x2pt analysis, will introduce a systematic bias of roughly 0.3\% bias on the auto-correlation, compared to a $0.5\%$ on both if the redshift evolution is simply neglected.  

The physical reason why a constant correction doesn't accurately account for the redshift dependence of the magnification bias is that the redshift dependence of the magnification kernel cannot be simply modelled by a change in amplitude. In \cref{appendix:redshift_dependence} we discuss this in more detail and compute the modification to the magnification kernel explicitly. We find that the redshift evolution of the sample is equivalent, to leading order, to a shift in the effective redshift of the sample, i.e. such that $\frac{d N}{d z} (z) \rightarrow \frac{d N}{d z} (z-\delta z_{\rm evo})$ when calculating the magnification kernel. For CMASS where the linear evolution is strongly detected we find $\delta z_{\rm evo} \approx 0.025$. This captures the redshift-dependent change to the magnification kernel and therefore is applicable to any observable, instead of for either the lensing cross-correlation or galaxy auto-correlation alone. To illustrate this we also show the correlation functions for CMASS using our baseline $\hat \alpha_{\rm CMASS}$ when including this shift in \cref{fig:cross_corr_redshift_evolution}.
Based on the freedom of the measured linear evolution, we can again estimate a systematic error on the magnification bias when applying this correction. We find a systematic error of $\pm 0.10$ for  $\hat \alpha_{\rm CMASS}$. More generally, one can also take into account the redshift evolution in a cosmological inference, especially for future surveys with larger constraining power, by sampling and marginalizing over the functional form of the redshift evolution (e.g. for a linear fitting form one would marginalize over the coefficients in \cref{eq:cmass_zdep}).

For LOWZ we do not find a clear linear evolution with redshift, but do find statistically significant differences when binning into four equal-weight subsamples. We linearly interpolate between our 4 bins to estimate the effect of the redshift dependence on observed spectra. We compare the resulting observables to the results when using the constant for the averaged sample. We find that for the cross-correlation with CMB lensing a shift approximately consistent with +0.06 and for the galaxy auto-correlation a shift of +0.07. Since this is a crude approach, we do not consider these as accurate enough to apply as a correction but rather consider them part of the systematic error budget. Therefore we estimate a systematic error for $\hat \alpha_{\rm LOWZ}$ of $\pm 0.07$ due to redshift evolution within the sample.

The z1 and z3 samples represent combinations of the CMASS and LOWZ datasets which have distinct magnification bias trends in redshift. We do not model their redshift dependence in this work but caution that we expect them to have significant redshift evolution that is not well described by a linear trend. Redshift evolution in the z1 and z3 samples is investigated in \citet{vonWietersheimKramsta2021} within their simulations-based approach.


\begin{table*}
\include{Tables/analysis_choices}
\caption{Overview for how different analysis choices impact the result for the magnification bias. For each only the statistical uncertainty is shown. \label{table:analysis_chocies}}
\end{table*}

\subsection{Impact of analysis choices on magnification bias}
\label{sec:impact_analysis_choices}

Finally, while in the main body of the paper we have been content to measure the magnification bias for the four LSS-oriented BOSS samples as defined in the original survey, in practice for cross-correlations analyses one often has to make various choices that slightly alter the sample composition. These choices include the sample restricted to certain areas of the sky or re-weighted in particular ways. In the case of BOSS, particularly relevant choices include North (NGC) versus South (SGC) galactic caps, which featured photometric targets observed using different telescopes, and weighting by FKP and cross-correlations weights (\cref{sec:SDSS_data}).

\cref{table:analysis_chocies} lists the magnification bias measured when accounting for these choices. The NGC and SGC samples show some weak ($2\sigma$) evidence for different magnification bias, particularly for the LOWZ sample; this is not entirely unexpected given the differences also in observed clustering between the sample \citep[see e.g. Fig.~7 of][]{Ross17}, and suggests that the two be treated as two samples with distinct (if similar) magnification biases in clustering analyses.

The most significant changes occur, however, before and after the application of redshift-dependent weights, particularly for the CMASS sample. Indeed, the FKP-weighted sample used for the fiducial BAO and RSD analyses of the BOSS collaboration \citep{Reid2016,Ross17}, and the cross-correlation weighted sample of \cite{Chen2022} differ by up to $18\sigma$\footnote{Here we consider only statistical uncertainties since they share systematic errors.}. This can be understood by noting that the weights applied have a significant effect on the mean redshift, and indeed interpolating to the new weighted mean redshift using \cref{eq:cmass_zdep} explains the majority of the difference in $\alpha$ between the various weight choices: for CMASS the FKP weights (cross-correlation weights) shift the weighted mean redshift up by 0.025 (down by 0.017) which from our linear trend predict changes of +0.22 (down by 0.15) in $\hat \alpha$, while for LOWZ the shifts are rather small.
Therefore the differences in average magnification bias we see for different analysis choices can be explained by the redshift dependence of $\alpha$ within the sample. This is another motivation to model the magnification bias in CMASS as a linear function in redshift as shown in \cref{eq:cmass_zdep} as this would reduce the dependence on analysis choices that change the effective redshift of the sample.

\section{Conclusions} 
\label{sec:conclusion}

Magnification bias, the contribution to the observed clustering of galaxies due to lensing from intervening matter, is a non-negligible signal in the angular auto and lensing cross-correlations of current and upcoming galaxy surveys. The effect on the lensing cross-correlation is particularly significant---of order $10\%$ for BOSS CMASS galaxies when cross-correlated with CMB lensing on large scales---leading to potential biases in 3x2pt weak lensing analyses if the effect is not properly modelled. 

The magnification bias of any sample can be split into two distinct contributions: a geometric contribution due to the apparent dilution of galaxies in the sky when magnified, and a survey-selection contribution $\alpha$ due to magnified galaxies dropping in or out of the selection criteria defining a given sample.

In this paper, we present a systematic accounting of the physical contributions to the latter, $\alpha$, (the size of the former is known to be $-2\kappa$) and propose an estimator based on artificially magnifying a galaxy sample by a series of external magnifications and fitting for the magnification bias using linear regression. Our estimator allows us to take advantage of reasonable step sizes with which to reduce the Poisson error on the magnification bias measurement while marginalizing out errors incurred by the finite difference assumed in simpler estimators (\cref{sec:mag_bias_derivation}). While our focus in this paper is on spectroscopic surveys, much of the same protocol can be applied to imaging surveys as well.

As an example, we measure the magnification bias for four samples in the BOSS survey---LOWZ and CMASS, as well as the merged redshift bins, z1 and z3. While we are not the first to investigate the magnification effect in this context \citep[see e.g.][]{Singh2019,vonWietersheimKramsta2021,Samuroff2023}, previous works have used estimates for the magnification bias that do not take into account the full gamut of physical effects lensing magnification has on the survey selection, and have either measured the logarithmic slope of the cumulative luminosity function at the edge of the survey selection following \cref{eq:log_slope} or calibrated the slope of the differential distribution to simulations. Our calculations demonstrate that the assumption of a magnitude-limited survey does not accurately reproduce the magnification bias for any of the BOSS samples as the most relevant photometric cuts are the \textit{colour-dependent} magnitude cuts along which most of these samples live. We also find that the lensing of galaxies needs to be carefully modelled, including the shape change under lensing. In \cref{sec:lensing_mags}, we derive in detail how the fiber and PSF magnitudes change under lensing. 
The cuts using these magnitudes make statistically relevant contributions for our samples, and not accounting for the shape change under lensing would significantly overestimate the effect of the corresponding photometric cuts on the magnification bias. In addition, we investigate the extent to which uncertainties in the precise shape profiles of galaxies affect our estimates and add the resulting systematic error to our budget. In future work, additional sources of systematic could be investigated, like uncertainties in the measurement of the light radius as well as measurement uncertainties of the observed magnitudes themselves. We can compare our results for the z1 and z3 samples with the results presented in \citet{vonWietersheimKramsta2021} where an empirical relation between the derivative of the differential magnitude distribution and the magnification bias was calibrated based on simulations. They found $\alpha = 2.62\pm0.28$ for the z3 sample which is around $1\sigma$ lower than our result, showing reasonable consistency. For the z1 sample, they report $\alpha = 1.93\pm 0.05$ which is significantly lower than our result. Their z1 sample is defined as $z>0.2$ whereas we use $z>0.15$  and therefore redshift evolution could explain the difference. Furthermore, their method relies on simulations that are representative of the data and they do report inaccuracies in the simulations at the low redshift end that could further explain the difference for the z1 result. 

We also investigated the LOWZ and CMASS samples for a redshift dependence of the magnification bias within the sample by binning them into subsamples. For LOWZ we only find a small dependence on redshift that can be characterized by a contribution to our systematic error budget. For CMASS we find a significant increase with redshift that needs to be accounted for in cosmological analyses. We discuss how the redshift-dependent change can not be accurately modelled as a constant correction to the averaged magnification bias. Instead, we discuss that a shift in the effective redshift of the sample can be used to account for the redshift dependence or one can sample over the linear function in redshift instead of a constant magnification bias in a cosmological analysis.   

Finally, we investigate a range of common analysis choices that can all affect the average magnification bias of the sample, especially when these choices change the effective redshift of the sample, such as weighting the sample by FKP weights for optimal power spectrum estimation or by an additional weight to match the effective redshifts of lensing cross-correlations to the galaxy auto-correlation. These redshift-dependent weights lead to detectable changes in the average magnification bias well beyond the statistical error for our sample, and care should therefore be taken to match magnification bias measurements specifically to one's analysis choices.

Other common approaches for the magnification bias can reduce the constraining power of cosmological analyses by either introducing potentially large systematic errors, like estimations based on simulations, or introducing large uncertainties by conservatively marginalizing the magnification bias with a wide prior. Estimating the magnification bias precisely and directly from data as presented in this work offers the opportunity to improve the accuracy with which the galaxy and lensing correlations can be connected to the underlying cosmology. This is critical to fully leverage the improved precision coming from imminent spectroscopic and photometric surveys since we expect them to have comparable levels of magnification bias.

%

\section*{Acknowledgments}
We thank Simone Ferraro, Martin White, and Rongpu Zhou for useful discussions that spurred the beginning of this project. We thank Alex Krolweski for helpful discussions about the role of redshift failures in the magnification bias. The work of L.W. and R.B. is supported by NSF grant AST-2206088, NASA ATP grant 80NSSC18K0695, and NASA ROSES grant 12-EUCLID12-0004. S.C.~is supported by the Bezos Membership at the Institute for Advanced Study.

Funding for SDSS-III has been provided by the Alfred P. Sloan Foundation, the Participating Institutions, the National Science Foundation, and the U.S. Department of Energy Office of Science. The SDSS-III web site is http://www.sdss3.org/.

SDSS-III is managed by the Astrophysical Research Consortium for the Participating Institutions of the SDSS-III Collaboration including the University of Arizona, the Brazilian Participation Group, Brookhaven National Laboratory, Carnegie Mellon University, University of Florida, the French Participation Group, the German Participation Group, Harvard University, the Instituto de Astrofisica de Canarias, the Michigan State/Notre Dame/JINA Participation Group, Johns Hopkins University, Lawrence Berkeley National Laboratory, Max Planck Institute for Astrophysics, Max Planck Institute for Extraterrestrial Physics, New Mexico State University, New York University, Ohio State University, Pennsylvania State University, University of Portsmouth, Princeton University, the Spanish Participation Group, University of Tokyo, University of Utah, Vanderbilt University, University of Virginia, University of Washington, and Yale University.

\section*{Data Availability}

The SDSS BOSS datasets used in the analysis are available under \url{https://data.sdss.org/sas/dr12/boss/lss/}. Our Python code to measure the magnification bias is available under \url{https://github.com/lukaswenzl/Magnification_bias_estimation_in_galaxy_surveys/}. This code can be applied to other samples by replacing the data and photometric selection specific to that survey.

\bibliographystyle{mnras}
\bibliography{main}

\appendix

\section{Redshift-Dependence of Magnification Bias} \label{appendix:redshift_dependence}

In \cref{sec:redshift_evolution} we have shown that the measurable redshift dependence of the LOWZ and especially CMASS sample introduces a systematic in the analysis when assuming a constant magnification bias for the whole sample. Especially interesting for the large evolution with redshift in CMASS is that the impact of the redshift dependence is different for different observables and can not exactly be modelled as a constant correction to the magnification bias $\alpha$. Physically this is because, unlike the case of shifting magnification bias due to different redshift weightings (\cref{sec:impact_analysis_choices}), which simply selects a new effective redshift for the magnification effect, the redshift-evolution effect comes about due to an interaction of the evolution of the lensing kernel and magnification bias. For example, in the case of CMASS, galaxies at higher redshift are both subject to more lensing \textit{and} more susceptible to it, leading to an enhancement quadratic in the redshift width of the galaxy distribution.

Let us try to understand this analytically. For a linearly evolving magnification bias (e.g. \cref{eq:cmass_zdep}) we can write the magnification kernel as $W^\mu = 2(\alpha_0-1) W^\mu_0 + 2\alpha_1 W^\mu_1$, where $\alpha_{0,1}$ are the zeroth and first derivatives of $\alpha(z)$ expanded about the mean redshift. Here we have defined
\begin{equation}
    W^\mu_0 = \int dz'\ \frac{dN}{dz'}\ W^{\kappa}(z,z') = W^\kappa(z,z_0) + \frac12 \left(\frac{d^2 W^\kappa}{dz'^2} \right)_0\Delta z^2 + ...,
\end{equation}
to be the magnification kernel for a constant $\alpha$, where $\Delta z^n$ is the nth moment of $dN/dz$ and $z_0$ its mean. Similarly
\begin{equation}
    W^\mu_1 = \int dz'\ \frac{dN}{dz'}\ (z' - z_0)  W^{\kappa}(z,z') = \left(\frac{d W^\kappa}{dz'} \right)_0 \Delta z^2 + ...,
\end{equation}
is the response of the magnification to a linear shift in magnification bias; evidently, the linear evolution only produces a change if the magnification kernel also evolves with source redshift. 
Note that the above integrals hold only for $z < z_{\rm min}$; for $z$ overlapping with the redshift range of the survey only galaxies in the background are lensed by matter at $z$, and for $z > z_{\rm max}$ the integral is zero. We then have
\begin{align}
    W^\mu(z) &= 2(\alpha_0-1) W^\kappa(z,z_0) \nonumber \\
    & \quad + \left[ (\alpha_0-1) \left(\frac{d^2 W^\kappa}{dz'^2} \right)_0 + 2\alpha_1\left(\frac{d W^\kappa}{dz'} \right)_0  \right] \Delta z^2 + ... \nonumber \\
    &= 2(\alpha_0-1) W^\kappa\left(z,z_0 + \frac{\alpha_1}{\alpha_0-1} \Delta z^2 \right) \nonumber \\
    & \quad  + (\alpha_0-1) \left(\frac{d^2 W^\kappa}{dz'^2} \right)_0 \Delta z^2 + ... \nonumber \\
    &\approx 2(\alpha_0 - 1) \int dz'\ \frac{dN}{dz''}(z''=z' - \frac{\alpha_1}{\alpha_0-1} \Delta z^2 )\ W^\kappa(z, z'), 
\end{align}
where the approximation is good to $\mathcal{O}(\Delta z^4)$, i.e. to leading order the redshift-evolution effect is equivalent to lensing the galaxies from a slightly higher redshift with the shift given by
\begin{equation}
    \delta z_{\rm evo} = \left(\frac{\alpha_1}{\alpha_0-1}\right) \Delta z^2.
\end{equation}
It is important to note that the derivation above only holds for a galaxy sample well-localized at a narrow peak around $z_0$, such that the magnification bias can be approximated as linearly evolving within the support of the galaxy distribution. For multiple-peaked galaxy distributions (e.g. LOWZ) the saddle-point approximation would pick out different shifts at each peak which suggests that such samples could be better modeled as a combination of different populations with different $dN/dz$'s.

\begin{figure}
\includegraphics[width=\columnwidth]{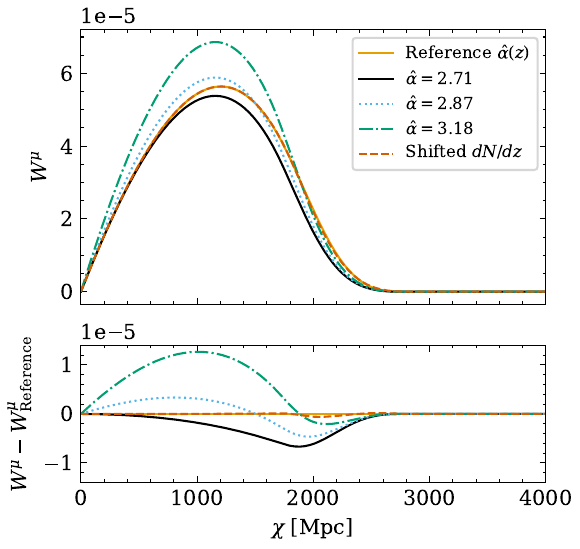}
\caption{Shown are the magnification kernels for CMASS with different magnification bias prescriptions. The upper panel shows the absolute values and the lower panel the difference to the magnification kernel for $\hat \alpha(z)$ [orange line]. A constant magnification bias even when shifted  [black line, dotted blue line, dash-dotted green line] can not capture the redshift-dependent change when including the redshift evolution. When instead shifting $dN/dz$ [dashed red line] we get good agreement with the magnification kernel for $\hat \alpha(z)$.}
\label{fig:magnification_kernels}
\end{figure}

\cref{fig:magnification_kernels} shows the magnification kernel for CMASS taking into account the linear redshift evolution of $\alpha(z)$ in \cref{eq:cmass_zdep}. The true magnification kernel, in this case, is not well-described by either the mean magnification bias [black line] or a shifted value to match the cross-correlation signal [blue dotted line]. The shifted value compensates for the underprediction at high redshifts by increasing the overall magnification bias but results in a redshift-dependent residual in $W^\mu$, and a scale-dependent residual in $\hat C_\ell^{\kappa g}$ (see \cref{fig:cross_corr_redshift_evolution}). In addition, the lensing cross-correlation and galaxy auto-correlation probe different scales and redshifts, leading to differently-shaped residuals when attempting to correct for one versus the other. We also show a shifted value to match the auto-correlation signal [dash-dotted green line] to highlight this. 
The red dashed line shows instead the correction obtained by shifting the galaxy distribution $dN/dz$ as described above. For CMASS we measure a shift of $\delta z_{\rm evo} \approx 0.025$, which corresponds well to the exactly-computed magnification kernel based on $\alpha(z)$ as shown.


\label{lastpage}
\end{document}

%% file: journaldef.tex
\def\aj{AJ}%
\def\apj{ApJ}%
\def\apjl{ApJ}%
\def\apjs{ApJS}%
\def\ao{Appl.~Opt.}%
\def\aap{A\&A}%
\def\mnras{MNRAS}%
\def\prd{Phys.~Rev.~D}%
\def\pasj{PASJ}%
\def\jcap{JCAP}%
\def\physrep{Phys.~Rep.}%

%% file: Tables/photometric_criteria.tex
\begin{tabular}{lll}
CMASS: $\hat \alpha_{\rm simple} = 2.70$ & Impact on $\hat \alpha_{\rm simple}$ \\ \hline
1. $d_{\perp}>0.55$ & 0 \\
2. $i_{\mathrm{cmod}}<19.86+1.6\left(d_{\perp }-0.8\right)$ & 1.68 \\
3. $17.5<i_{\mathrm{cmod}}<19.9$ & 0.90 \\
4. $r_{\mathrm{mod}}-i_{\mathrm{mod}}<2$ & 0 \\
5. $i_{\mathrm{fib} 2}<21.5$ & 0.26 \\
6. $i_{\mathrm{psf}}-i_{\mathrm{mod}}>0.2+0.2\left(20.0-i_{\mathrm{mod}}\right)$ & 0.01 \\
7. $z_{\mathrm{psf}}-z_{\mathrm{mod}}>9.125-0.46 z_{\mathrm{mod}}$ & -0.05 \\ \hline
LOWZ: $\hat \alpha_{\rm simple} = 2.43$  \\ \hline
1. $\left|c_{\perp}\right|<0.2$ & 0 \\
2. $r_{\mathrm{cmod}}<13.5+c_{\parallel} / 0.3$ & 2.16 \\
3. $16<r_{\mathrm{cmod}}<19.6$ & 0.30 \\
4. $r_{\mathrm{psf}}-r_{\mathrm{cmod}}>0.3$ & 0.00 
\end{tabular}

%% file: Tables/baseline_results.tex
\begin{tabular}{lllllllll}
 & Estimate & Statistical Uncertainty & Light profile choice \\\hline
$\hat \alpha_{\rm CMASS}$ & $2.71$ & $\pm 0.02$ & $\pm 0.02$ \\
$\hat \alpha_{\rm LOWZ}$ & $2.45$ & $\pm 0.02$ & $\pm 0.00$ \\
$\hat \alpha_{\rm z3}$ & $2.93$ & $\pm 0.02$ & $\pm 0.06$ \\
$\hat \alpha_{\rm z1}$ & $2.35$ & $\pm 0.02$ & $\pm 0.06$ \\
\end{tabular}

%% file: Tables/analysis_choices.tex
\begin{tabular}{lllll}
 Analysis choices & \multicolumn{4}{c}{$\hat \alpha$ for} \\
 & CMASS & LOWZ & z3 & z1 \\ \hline
Baseline & $2.71\pm 0.02$ & $2.45\pm 0.02$ & $2.93\pm 0.02$ & $2.35\pm 0.02$ \\
North sample only & $2.73 \pm 0.02$ & $2.49 \pm 0.03$ & $2.95 \pm 0.02$ & $2.37 \pm 0.02$ \\
South sample only & $2.67 \pm 0.03$ & $2.36 \pm 0.04$ & $2.90 \pm 0.04$ & $2.31 \pm 0.03$ \\
No redshift cuts & $2.74 \pm 0.02$ & $2.36 \pm 0.02$ & - & - \\
Adding FKP weights & $3.00 \pm 0.02$ & $2.48 \pm 0.02$ & $3.24 \pm 0.03$ & $2.37 \pm 0.02$ \\
Adding reweighting for cross-corr. & $2.51 \pm 0.02$ & $2.43 \pm 0.02$ & $2.68 \pm 0.02$ & $2.32 \pm 0.02$ \\
Adding FKP weights and reweighting for cross-corr. & $2.63 \pm 0.02$ & $2.45 \pm 0.02$ & $2.82 \pm 0.02$ & $2.33 \pm 0.02$ 
\end{tabular}